\documentclass[aps,prl,twocolumn,superscriptaddress,showpacs,amsmath,amssymb]{revtex4}

\usepackage{graphicx}

\begin{document}

\title{Many-body quantum ratchet in a Bose-Einstein condensate}

\author{Dario Poletti}

\affiliation{Department of Physics and Centre for Computational Science and Engineering,
National University of Singapore, Singapore 117542,
Republic of Singapore}

\author{Giuliano Benenti}

\affiliation{Center for Nonlinear and Complex Systems, Universit\`a degli
Studi dell'Insubria, Via Valleggio 11, 22100 Como, Italy}
\affiliation{Istituto Nazionale di Fisica Nucleare, Sezione di Milano,
CNISM and CNR-INFM}

\author{Giulio Casati}

\affiliation{Center for Nonlinear and Complex Systems, Universit\`a degli
Studi dell'Insubria, Via Valleggio 11, 22100 Como, Italy}
\affiliation{Istituto Nazionale di Fisica Nucleare, Sezione di Milano,
CNISM and CNR-INFM}
\affiliation{Department of Physics and Centre for Computational Science and Engineering,
National University of Singapore, Singapore 117542,
Republic of Singapore}

\author{Baowen Li}

\affiliation{Department of Physics and Centre for Computational Science and Engineering,
National University of Singapore, Singapore 117542,
Republic of Singapore}
\affiliation{Laboratory of Modern Acoustics and Institute of Acoustics,  Nanjing University, 210093, P R China}
\affiliation{NUS Graduate School for Integrative Sciences and Engineering, 117597, Republic of Singapore}

\date{\today}

\begin{abstract}
We study the dynamics of a dilute Bose-Einstein condensate confined in a toroidal trap
and exposed to a pair of periodically flashed optical lattices. We first prove that in the noninteracting case
this system can present a quantum symmetry which forbids the ratchet effect classically expected.
We then show how many-body atom-atom interactions, treated within the mean-field approximation, can break this
quantum symmetry, thus generating directed transport.
\end{abstract}



\maketitle

The ratchet effect, that is, the possibility to drive directed transport with the help
of zero-mean perturbations, has recently gained renewed attention due to its possible
relevance for biological transport, molecular motors and the prospects
of nanotechnology \cite{Hanggi,Reimann}.
At the classical level, the ratchet effect can be found in periodic systems
due to a broken space-time symmetry \cite{Flach}.
The ratchet phenomenon has also been discussed in quantum systems \cite{Hanggi2},
including the Hamiltonian limit without dissipation \cite{Monteiro}.
Experimental implementations of directed transport range from semiconductor heterostructures
to quantum dots, Josephson junctions, and cold atoms in optical lattices \cite{Renzoni}.

Quantum Hamiltonian ratchets are relevant in systems such as cold atoms in which
the high degree of quantum control may allow experimental implementations
near to the dissipationless limit. Moreover, the realization of Bose-Einstein condensates (BECs)
of dilute gases has opened new opportunities for the study of dynamical systems
in the presence of many-body interactions. Indeed, it is possible to prepare
initial states with high precision and to tune over a wide range the
many-body atom-atom interaction.
From the viewpoint of directed transport, the study of
many-body quantum system is, to our knowledge, at the very beginning.

In this Letter, we investigate the quantum dynamics of a
BEC in a pair of periodically flashed optical lattices.
We show how the interaction between atoms in the condensate, studied in the mean-field approximation,
can break the quantum symmetry present in our model in the noninteracting limit,
thus giving rise to the ratchet effect.
The role of noise, the validity of the mean-field description and the possibility to observe experimentally our
ratchet model are discussed as well.

%
%

We consider $N$ condensed atoms confined in a toroidal trap of radius $R$ and cross
section $\pi r^2$, with the condition $r\ll R$, so that the motion is essentially
one-dimensional. The dynamics of a dilute condensate in a pair of periodically
kicked optical lattices at zero temperature is described by the Gross-Pitaevskii
nonlinear equation,
\begin{equation}
i\frac{\partial}{\partial t}\psi(\theta,t)=
\left[-\frac{1}{2}\frac{\partial^2}{\partial \theta^2} +
g|\psi(\theta,t)|^2 + V(\theta,\phi,t)\right]\psi(\theta,t),
\label{eq:grosspitaevskii}
\end{equation}
where $\theta$ is the azimuthal angle,
$g=8NaR/r^2$ is the scaled strength of the nonlinear interaction
(we consider the repulsive case, \emph{i.e.}, $g>0$),
$a$ is the $s$-wave scattering length
for elastic atom-atom collisions.
The kicked potential $V(\theta,\phi,t)$ is defined as
\begin{equation}
\begin{array}{c}
V(\theta,\phi,t)=\sum_n[V_1(\theta)\delta(t-nT)+V_2(\theta,\phi)\delta(t-nT-\xi)],
\\
\\
V_1(\theta)=k\cos\theta,\;\;
V_2(\theta,\phi)=k\cos(\theta-\phi),
\end{array}\label{eq:potential}
\end{equation}
where $k$ is the kicking strength and $T$ the period of the kicks.
The parameters $\phi\in [0,2\pi]$ and $\xi\in[0,T]$ are used to
break the space and time symmetries, respectively. Note that we
set $\hbar=1$ and that the length and the energy are measured in
units of $R$ and $\hbar^2/m R^2$, with $m$ the atomic mass. The
wave function normalization reads $\int_0^{2\pi} d\theta
|\psi(\theta,t)|^2=1$ and boundary conditions are periodic,
$\psi(\theta+2\pi,t)=\psi(\theta,t)$.

%
%

We first consider the noninteracting case $g=0$. Here, when $\phi\ne 0,\pi$ and $\xi\ne 0, T/2$
space-time symmetries are broken and there is directed transport, both in the classical limit and,
in general, in quantum mechanics \cite{Gabriel2}.
However, if we take $T=6\pi$ and $\xi=4\pi$, then the quantum motion,
independently of the kicking strength $k$,
is periodic of period $2T$.

In order to prove this periodicity, it is useful to write the initial
wave function as
$\psi(\theta,0)=\sum_n A_n\exp(in\theta)$, where
$A_n=\frac 1 {2\pi}\int_0^{2\pi}\psi(\theta,0)\exp(-in\theta)$.
After free evolution up to time $t$, the wave function becomes $\psi(\theta,t)=\sum_n
A_n\exp\left(-i\frac{n^2}{2}t +in\theta\right)$. If $t=4\pi$ we
have $\psi(\theta,t)=\psi(\theta,0)$ while, if $t=2\pi$, we obtain
$\psi(\theta,t)=\psi(\theta+\pi,0)$.
Using these relations we can easily see that the system is periodic with
period $12\pi$. Indeed,
\begin{equation}
\begin{array}{c}
\psi(\theta,4\pi^+)=\exp[-iV_1(\theta)]\psi(\theta,0),\\
\\
\psi(\theta,6\pi^+)=\exp[-iV_2(\theta,\phi)]\psi(\theta+\pi,4\pi^+)\\
=\exp\{-i[V_2(\theta,\phi)-V_1(\theta)]\}\psi(\theta+\pi,0),\\
\\
\psi(\theta,10\pi^+)=\exp[-iV_1(\theta)]\psi(\theta,6\pi^+)\\
=\exp(-iV_2(\theta,\phi))\psi(\theta+\pi,0),\\
\\
\psi(\theta,12\pi^+)=\exp[-iV_2(\theta,\phi)]\psi(\theta+\pi,10\pi^+)=\psi(\theta,0),
\end{array}
\end{equation}
where $\psi(\theta,t^+)$ denotes the value of the
wave function at time $t$ just after the kick.
The momentum
$\langle p(t) \rangle = -i \int_0^{2\pi} d\theta \psi^\star(\theta,t)
\frac{\partial}{\partial \theta} \psi(\theta,t)$ also changes
periodically with period $12\pi$ (4 kicks).
Therefore, the average momentum
$p_{\rm av}=\lim_{t\to\infty} \overline{ p }(t)$
($\overline{ p }(t)=\frac{1}{t}\int_0^t dt^\prime
\langle p(t^\prime) \rangle$)
is given by
\begin{equation}
\begin{array}c
p_{\rm av}=\frac{4\pi\langle p(0)\rangle + 2\pi\langle p(4\pi^+)\rangle +
4\pi\langle p(6\pi^+)\rangle +2\pi\langle p(10\pi^+)\rangle}{12\pi}\\
\\
=\langle p(0)\rangle+\frac k 2 \int_0^{2\pi}
\left( \sin(\theta) - \sin(\theta-\phi) \right)|\psi(\theta,0)|^2d\theta.
\end{array}
\end{equation}
In particular, for the constant initial condition
$\psi(\theta,0)=1/\sqrt{2\pi}$, which is the ground state of a
particle in the trap, the momentum remains zero at any later time.
This initial condition has an important physical meaning, as it
corresponds to the initial condition for a Bose-Einstein
condensate.

%
%

\begin{figure}[!h]
\includegraphics[width=\columnwidth]{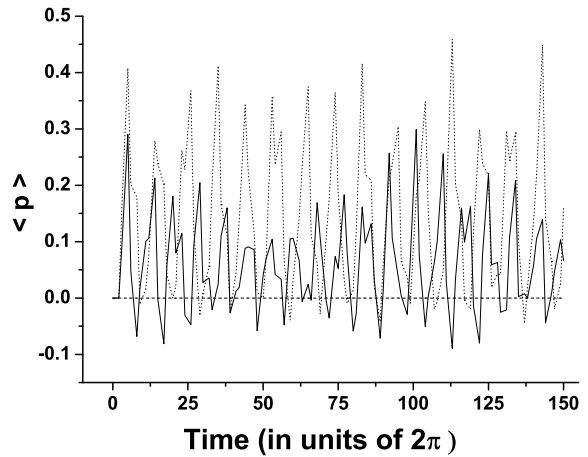}
\vspace{-0.6cm} \caption{Momentum versus time for different values of
interaction strength $g$, at $k\approx 0.74$ and $\phi=-\pi/4$: $g=0$ (dashed line),
$g=0.5$ (continuous curve), $g=1$ (dotted curve).}
\label{fig:momg}
\end{figure}

It is therefore interesting to study the case of a BEC because
atom-atom interactions may break the above periodicity,
and this may cause generation of momentum.
The numerical integration of Eq.~(\ref{eq:grosspitaevskii})
confirms this expectation:
as shown in Fig.~\ref{fig:momg}, at $g\ne 0$ the momentum oscillates around a mean
value clearly different from zero. Notice that without interactions
($g=0$) the momentum is exactly zero, so that directed transport is induced
by the many-body atom-atom interactions.

%
%

\begin{figure}[!h]
\includegraphics[width=\columnwidth]{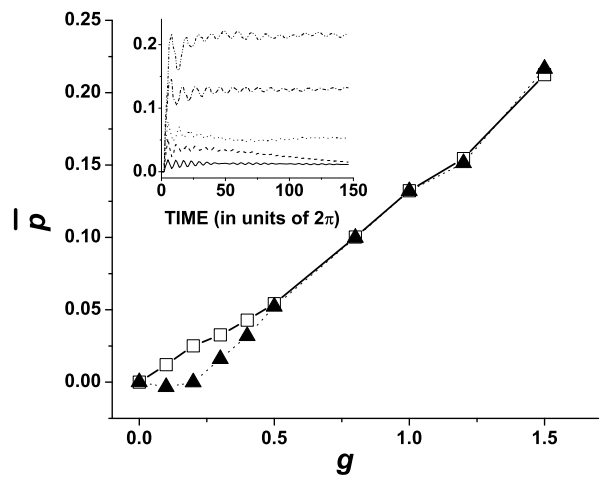}
\vspace{-0.6cm} \caption{Momentum averaged over the first $30$ kicks
(solid line with boxes) and asymptotic momentum
(dotted line with triangles). Inset: Cumulative average $\overline{ p }(t)$
as a function of time for different values of $g$.
From bottom to top $g=0.1, 0.2, 0.4, 1.0, 1.5$.
Parameter values: $k\approx 0.74$, $\phi=-\pi/4$.}
\label{fig:pg}
\end{figure}

In Fig.~\ref{fig:pg}, we compare the asymptotic value $p_{\rm av}$, obtained from long numerical integrations of the Gross-Pitaevskii
equation (dotted line with triangles), with the average of $\langle p(t) \rangle$ over the first $30$ kicks ($\bar{p}(90\pi)$, continuous line with boxes).
It can be seen that this short-time average is sufficient to obtain a good estimate of
the average momentum $p_{\rm av}$, provided that $g\gtrsim 0.5$
It is interesting to remark that the average momentum after the first kicks grows monotonously with $g$.
Therefore, the ratchet current provides a method to measure the interaction strength in an
experiment.
In the inset of Fig.~\ref{fig:pg} we show the cumulative average
$\overline{ p }(t)$.
For strong enough interactions
($g\gtrsim 0.5$)
the convergence to the limiting value $p_{\rm av}$
is rather fast as we can already see from the main part of Fig.~\ref{fig:pg}.

%
%

Since the above complete periodicity of the single particle system ($g=0$) is a
very fragile quantum phenomenon, it is important to check the visibility of
the ratchet effect when the unavoidable noise leads to a departure from
the ideal periodic behavior. For this purpose we consider fluctuations in the kicking
period, modeled as random and memoryless variations of the period between consecutive kicks,
with the fluctuation amplitude at each kick randomly drawn from a uniform distribution in
the interval $[-\epsilon, \epsilon]$. We have seen that, when the
the size of the fluctuations is $\epsilon=T/200$, then the ratchet
current generated in the noninteracting case is $\bar{p}(90\pi)=-0.007$. This value of $\bar{p}$ is much smaller than the
genuine many-body ratchet current already shown in Fig.~\ref{fig:pg} for values of $g\gtrsim 0.2$. A similar conclusion is obtained when the kicks are substituted by more
realistic gaussian pulses of width $T/10$. In this case $\bar{p}$ after
$30$ kicks is equal to $\bar{p}(90\pi)=-0.01$, again too small to hide or to be confused with the many-body
ratchet effect.

%
%

The interaction-induced generation of a nonzero current can be understood as follows.
We approximate, for small values of $g$, the free evolution of the BEC by a
split-operator method as in \cite{Dario1}:
\begin{equation}
\psi(\theta,\tau)\approx
e^{-i \frac{1}{2} \frac{\partial^2}{\partial \theta^2}\frac{\tau}{2}}
e^{-i g\left|\tilde{\psi}\left(\theta,\frac{\tau}{2}\right)\right|^2 \tau}
e^{-i \frac{1}{2} \frac{\partial^2}{\partial \theta^2}\frac{\tau}{2}}
\psi(\theta,0),
\label{eq:splitoperator}
\end{equation}
where $\tilde{\psi}\left(\theta,t+\Delta t\right)=e^{-i \frac{1}{2} \frac{\partial^2}{\partial \theta^2}\Delta t}
\psi(\theta,t)$.

In particular, we obtain
\begin{equation}
\psi(\theta,4\pi)\approx\exp(-i 4\pi g |\psi(\theta,0)|^2)\psi(\theta,0)\label{eq:evol4}
\end{equation}
and
\begin{equation}
\begin{array}c
\psi(\theta,2\pi)\approx\exp\{-i \pi g [|\psi(\theta,0)|^2+|\psi(\theta+\pi,0)|^2]\}\\
\times \{\cos[F(\theta,0) ]\psi(\theta+\pi,0) -\sin[F(\theta,0)] \psi(\theta,0)\},
\end{array}\label{eq:evol2}
\end{equation}
where we have defined
$F(\theta,0)= i \pi g [\psi^*(\theta,0)\psi(\theta+\pi,0) -\psi(\theta,0)\psi^*(\theta+\pi,0)]$.
Note that, in the limit  $g\rightarrow 0$, Eqs.~(\ref{eq:evol4}) and (\ref{eq:evol2})
become $\psi(\theta,4\pi)=\psi(\theta,0)$ and
$\psi(\theta,2\pi)=\psi(\theta+\pi,0)$, as expected for the noninteracting free evolution.
Using this approximation, we compute the evolution of the condensate for the first two kicks,
starting from the initial condition $\psi(\theta,0)=1/\sqrt{2\pi}$. We obtain
\begin{equation}
\begin{array}c
\psi(\theta,4\pi^+)\approx \frac{1}{\sqrt{2\pi}}\exp[-iV_1(\theta)]\exp(-i 2g),\\
\\
\psi(\theta,6\pi^+)\approx \frac{1}{\sqrt{2\pi}}\exp\{-i[V_2(\theta,\phi)-V_1(\theta)]\}
\exp(-i 3g)\\
\times\left\{ \cos[\Omega_1(\theta)] + \sin[\Omega_1(\theta)]\exp[-i2V_1(\theta)]\right\},
\end{array}\label{eq:after}
\end{equation}
where $\Omega_1(\theta)=g\sin(2V_1(\theta))$.
The mechanism of the ratchet effect is now clear:
due to atom-atom interactions, the modulus square of the wave function at time $6\pi$
(before the second kick) is no longer constant in $\theta$.
Instead we have, to first order in $g$,
$|\psi(\theta,6\pi)|^2\approx \frac{1}{2\pi}\{1+g\sin[4V_1(\theta)]\}$,
so that the initial constant probability distribution is
modified by a term symmetric under the transformation $\theta\to -\theta$.
The current after the kick at time $t=6\pi$ is then given by
\begin{equation}
\begin{array}{c}
\langle p (6\pi^+) \rangle = -\int_0^{2\pi} d\theta V_2^\prime (\theta,\phi) |\psi(\theta,6\pi)|^2
\\
\approx gk \int_0^{2\pi} d\theta \sin(\theta-\phi) \sin(4k\cos\theta)
=-gk\sin(\phi)J_1(4k),
\end{array}
\label{eq:perturbativecurrent}
\end{equation}
where $J_1$ is the Bessel function of the first kind of index $1$.
This current is in general different from zero, provided that
$V_2(\theta,\phi)$ is not itself symmetric under $\theta\to -\theta$,
that is, when $\phi\ne 0,\pi$.

%
%

\begin{figure}[!h]
\includegraphics[width=\columnwidth]{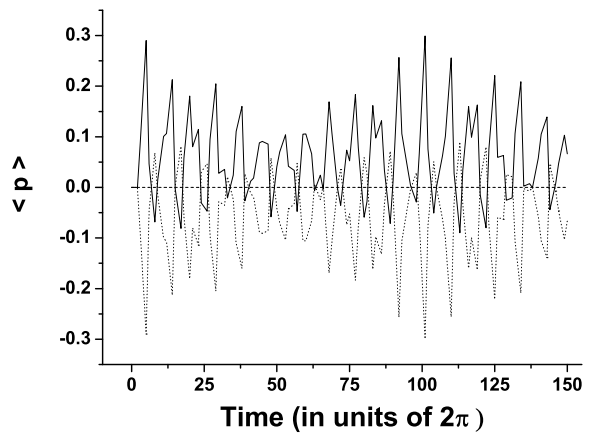}
\vspace{-0.6cm} \caption{Momentum versus time for different values of
the parameter $\phi$, at $k\approx 0.74$ and $g=0.5$:
$\phi=-\pi/4$ (continuous curve), $\phi=0$ (dashed line), $\phi=\pi/4$ (dotted curve).}
\label{fig:g0p5diffa}
\end{figure}

In Fig.~\ref{fig:g0p5diffa}, we show that it is possible to control the direction of
transport by varying the phase $\phi$: the current can be reversed simply by changing
$\phi\to -\phi$.
This current inversion can be explained by means of the following symmetry considerations.
The evolution of the wave-function $\psi(\theta,t)$ is given by Eq.~(\ref{eq:grosspitaevskii}).
After substituting in this equation $\theta\rightarrow-\theta$, and taking into account that that
$V(-\theta,\phi,t)=V(\theta,-\phi,t)$, we obtain
\begin{equation}
i\frac{\partial}{\partial t}\tilde{\psi}(\theta,t)=
\left[-\frac{1}{2}\frac{\partial^2}{\partial \theta^2} +
g|\tilde{\psi}(\theta,t)|^2 + V(\theta,-\phi,t)\right]\tilde{\psi}(\theta,t),
\end{equation}
where $\tilde{\psi}(\theta,t)\equiv \psi(-\theta,t)$.
Therefore, if $\psi(\theta,t)$ is a solution of the Gross-Pitaevskii equation, then also
$\tilde{\psi}(\theta,t)$ is a solution, provided that we substitute $\phi\rightarrow -\phi$ in the
potential $V$.
The momentum $\langle \tilde{p}(t)\rangle$ of the wavefunction $\tilde{\psi}(\theta,t)$ is obviously given
by $\langle \tilde{p}(t)\rangle=-\langle p(t)\rangle$, where $\langle p(t)\rangle$ is the
momentum of $\psi(\theta,t)$.
This means that, for every $\psi(\theta,t)$ whose evolution is ruled by the Gross-Pitaevskii equation
with potential $V(\theta,\phi,t)$, the wavefunction
$\tilde{\psi}(\theta,t)$ evolves with exactly opposite momentum if
$\phi\rightarrow -\phi$ in $V$.
Since we start with an even wavefunction, $\tilde{\psi}(\theta,0)=\psi(-\theta,0)=\psi(\theta,0)$,
then changing $\phi\rightarrow-\phi$ changes the sign of the momentum of the wavefunction at any later time.

%
%

When studying the dynamics of a kicked BEC, it is important to
take into account the proliferation of noncondensed atoms.
Actually, strong kicks may lead to
thermal excitations out of equilibrium and destroy the condensate, rendering
the description by the Gross-Pitaevskii equation meaningless
\cite{zoller,Jie}.
In the following, we show that, for the parameter values considered in this paper,
the number of noncondensed particles is negligible compared to the number of condensed ones,
thus demonstrating that
our theoretical and numerical results
based on the Gross-Pitaevskii equation are reliable.

Following the approach developed in \cite{Castin}
(see also \cite{Jie}), we compute the mean number of noncondensed particles
at zero temperature as $\delta N(t)=\sum_{j=1}^{\infty} \int_0^{2\pi}
d\theta |v_j(\theta,t)|^2$, where the
evolution of $v_j(\theta,t)$ is determined by
\begin{equation}
i\frac {\partial}{\partial t}\left[\begin{array}c
u_j(\theta,t)\\
v_j(\theta,t)
\end{array}\right]=\left[\begin{array}{cc}
H_1(\theta,t) & H_2(\theta,t)\\
-H_2^*(\theta,t) & -H_1^\star(\theta,t)
\end{array}\right]\left[\begin{array}c
u_j(\theta,t)\\
v_j(\theta,t)
\end{array}\right].
\label{eq:noncondensed}
\end{equation}
Here $H_1(\theta,t)=H(\theta,t)-\mu(t)+
g{Q}(t)|\psi(\theta,t)|^2{Q}(t)$,
$H(\theta,t)= -\frac 1 2 \frac{\partial^2}{\partial\theta^2}+
g|\psi(\theta,t)|^2+V(\theta,\phi,t)$ is the
mean-field Hamiltonian that governs the Gross-Pitaevskii equation
(\ref{eq:grosspitaevskii}), $\mu(t)$ is the chemical
potential ($H(\theta,t)\psi(\theta,t)=\mu(t)\psi(\theta,t)$),
${Q}(t)=\openone-|\psi(t)\rangle\langle\psi(t)|$
projects orthogonally to $|\psi(t)\rangle$,
and $H_2(\theta,t)=g{Q}(t)\psi^2(\theta,t){Q}^*(t)$.

We integrate in parallel Eqs.~(\ref{eq:grosspitaevskii}) and (\ref{eq:noncondensed}).
The initial condition of the noncondensed part is obtained by diagonalizing
the linear operator in (\ref{eq:noncondensed}) \cite{Castin,Jie}.
We obtain
\begin{equation}
\left(\begin{array}c
u_j(\theta,0)\\
v_j(\theta,0)\\
\end{array}\right)=\frac 1 2 \left(\begin{array}c
\xi+1/\xi\\
\xi-1/\xi\\
\end{array}\right) \frac{e^{ij\theta}}{\sqrt{2\pi}},
\end{equation}
where $\xi=\left(\frac{j^2/2}{j^2/2+2g|\psi(\theta,0)|}\right)^{1/4}$, with
$\psi(\theta,0)=1/\sqrt{2\pi}$ initial condition of the BEC.
The numerical evolution is performed using the split-operator method
as in Eq.~(\ref{eq:splitoperator}), with small integration steps
$\tau\ll T$.

\begin{figure}[!h]
\includegraphics[width=\columnwidth]{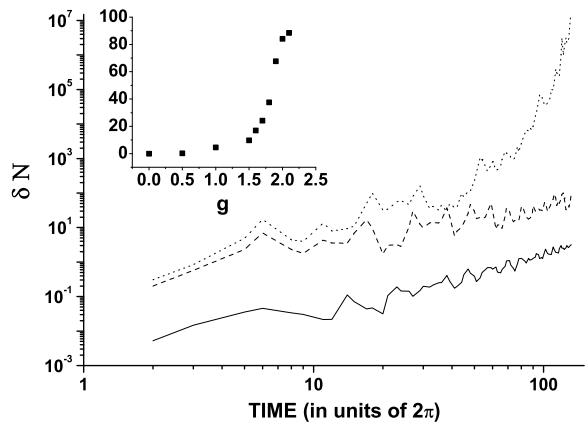}
\vspace{-0.6cm} \caption{Mean number $\delta N$ of noncondensed particles versus time for
different values of the interaction strength $g$:
from bottom to top, $g=0.5$, $1.5$, and $2.0$.
Inset: $\delta N$ vs. $g$ after $30$ kicks.
Parameter values: $k\approx 0.74$, $\phi=-\pi/4$.}
\label{fig:stabilita}
\end{figure}

The number $\delta N$ of noncondensed particles, depending on the stability or instability
of the condensate, grows polynomially or exponentially.
As shown in Fig.~\ref{fig:stabilita}, $\delta N$
grows polynomially at small $g$ and exponentially for large $g$.
The transition from stability
to instability takes place at $g=g_c\approx 1.7$. At $g>g_c$, thermal
particles proliferate exponentially fast, $\delta N \sim \exp(rt)$,
leading to a significant depletion of the condensate after a time
$t_d\sim \ln(N)/r$.
On the other hand, for $g<g_c$
the exponential growth rate $r=0$ and the number of noncondensed particles
is negligible for up to long times.
For instance, as shown in Fig.~\ref{fig:stabilita},
$\delta N \approx 0.2\;(10)$ after $t=90\pi$ (30 kicks) at $g=0.5\;(1.5)$,
which is much smaller than the total number of particles $N\approx 10^3-10^5$ \cite{Jie,Raizen}.

%
%

Finally, we would like to comment on the experimental feasibility of our proposal.
The torus-like potential confining the BEC may be realized by means of
optical billiards \cite{Milner}. The kicks may be applied using a
periodically pulsed strongly detuned laser beam with a suitably engineered
intensity, as proposed in \cite{Graham}.
The feasibility is also supported by the latest progresses in the
realization of BECs in optical traps such as the ${}^{87}$Rb BEC
in a quasi-one-dimensional optical box trap, with condensate length $\sim 80$ $\mu$m, transverse
confinement $\sim 5$ $\mu$m, and number of particles $N\sim 10^3$ \cite{Raizen}.
Sequences of up to $25$ kicks have been applied to a BEC of ${}^{87}$Rb atoms
confined in a static harmonic magnetic trap, with kicking strength $k\sim 1$
and in the quantum antiresonance case for the kicked oscillator model,
$T=2\pi$ \cite{Duffy}.
Finally, the interaction strength $g$ can be tuned over a very large range
using a Feshbach resonance \cite{Donley}.

This work was supported in part by the MIUR COFIN-2005.

\end{document}